\documentclass{elsart}
\usepackage{amsmath,amssymb,amsfonts}
\usepackage{graphicx,subfigure}
\usepackage{cite}

\renewcommand{\det}{\text{det}}

\renewcommand{\l}{\left}
\renewcommand{\r}{\right}
\begin{document}
\begin{frontmatter}
\title{The Antonov problem for rotating systems}

\author{A. De Martino,}
\author{E.V. Votyakov}
\author{D.H.E. Gross}
\address{Hahn-Meitner-Institut, Bereich Theoretische Physik\\
Glienickerstr. 100, 14109 Berlin (Germany)}
\date{}
\begin{abstract}
We study the classical Antonov problem (of retrieving the
statistical equilibrium properties of a self-gravitating gas of
classical particles obeying Boltzmann statistics in space and
confined in a spherical box) for a rotating system. It is shown
that a critical angular momentum $\lambda_c$ (or, in the canonical
language, a critical angular velocity $\omega_c$) exists, such
that for $\lambda<\lambda_c$ the system's behaviour is
qualitatively similar to that of a non-rotating gas, with a high
energy disordered phase and a low energy collapsed phase ending
with Antonov's limit, below which there is no equilibrium state.
For $\lambda>\lambda_c$, instead, the low-energy phase is
characterized by the formation of two dense clusters (a ``binary
star''). Remarkably, no Antonov limit is found for
$\lambda>\lambda_c$. The thermodynamics of the system (phase
diagram, caloric curves, local stability) is analyzed and compared
with the recently-obtained picture emerging from a different type
of statistics which forbids particle overlapping.
\end{abstract}
\end{frontmatter}

\section{Introduction}

The name ``Antonov problem'' is usually referred to the study of
the mean-field equilibrium state of self-gravitating $N$-body
systems of classical particles in a finite volume. One considers
the system with Hamiltonian
\begin{gather} H_N\equiv
H_N(\{\boldsymbol{r}_i\},\{\boldsymbol{p}_i\})=\frac{1}{2}
\sum_{i=1}^N p_i^2+\Phi(\{\boldsymbol{r}_i\})\label{ham}\\
\Phi(\{\boldsymbol{r}_i\})=-G\sum_{1\leq i<j\leq N}
\frac{1}{|\boldsymbol{r}_i-\boldsymbol{r}_j|}\label{pot}
\end{gather}
enclosed in a finite three-dimensional spherical box of radius
$R$, which is necessary in order to prevent particles from
evaporating. In (\ref{ham},\ref{pot}): $\boldsymbol{r}_i\in
V\subset\mathbb{R}^3$ and $\boldsymbol{p}_i\in\mathbb{R}^3$
denote, respectively, the position and momentum of the $i$-th
particle, $G$ is the gravitational constant, and $V$ stands for
the (volume of the) box containing the particles; we set the
particles' masses to one. The main problems are: (a) find the
equilibrium configurations, i.e. the particle density profiles
$\rho(\boldsymbol{r})$
($\int_V\rho(\boldsymbol{r})d\boldsymbol{r}=N$) that maximize the
microcanonical entropy ($\alpha=$ constant)
\begin{equation}
S_N\equiv S_N(E)=\log\l[\frac{\alpha}{N!}\int \delta(H_N-E)
D\boldsymbol{r}D\boldsymbol{p}\r]
\end{equation}
($D\boldsymbol{x}=\prod_{i=1}^N d\boldsymbol{x}_i$); and (b)
analyze the corresponding thermodynamics. Clearly, the existence
of such profiles and their actual structure will depend on the
value of the total energy $E$. Intuitively, when $E$ is large and
positive (that is when the kinetic term dominates) a gaseous-type
of structure will be preferred, with particles uniformly spread
over $V$, whereas when $E$ is sufficiently negative particles are
expected to attract each other in a dense core embedded in a
vapor. The situation presents however several subtleties.
\emph{Assuming that particles are point masses obeying Boltzmann
statistics in space}, i.e. that they can come arbitrarily close to
each other, it is found within the mean-field approximation
\cite{antonov62,lyndenbell68,padmanabhan89,padmanabhan90,chavanis01}
that
\begin{enumerate}
\item[(a)] there is no global entropy maximum, therefore all maxima
are local and correspond to metastable situations;
\item[(b)] if $\frac{RE}{GN^2}\equiv\epsilon>\epsilon_c\simeq -0.335$
(i.e. at sufficiently high energies, for we take $R$ to be fixed)
there exist local entropy extrema where the particle density
satisfies the condition
\begin{equation}\label{is}
\rho(\boldsymbol{r})=A\exp\l[\beta
\int\frac{\rho(\boldsymbol{r'})}{
|\boldsymbol{r}-\boldsymbol{r'}|}d\boldsymbol{r'}\r]
\qquad\beta=\frac{3/2}{\l[E-\Phi(\{\boldsymbol{r}_i\})\r]}
\end{equation}
where $A$ is a constant. Notice that $\beta$ is itself a function
of $\rho$. The physically admissible solutions of (\ref{is}) are
spherically symmetric (in fact, they can be mapped onto the
solutions of the Emden equation $y''+2y'/x=e^{-y}$, with
$y(0)=y'(0)=0$) and are called microcanonical ``isothermal
spheres'';
\item[(c)] in general, local extrema in the space of $\rho$'s
can be maxima (metastable states) or saddle points (unstable).
Microcanonical isothermal spheres are local maxima if the
``density contrast'' $\rho(0)/\rho(R)\lesssim 709$, otherwise they
are at most saddle points of the entropy surface;
\item[(d)] if $\epsilon$ lies in the range between $-0.2$ and
$\epsilon_c$ isothermal spheres have negative specific
heat\footnote{As first pointed out by Thirring \cite{thirring70},
in the microcanonical ensemble self-gravitating systems can have
negative specific heat, at odds with the canonical setup, in which
it is related to energy fluctuations and positive definite.}: in
such states the system heats up and shrinks upon decreasing
energy;
\item[(e)] finally, if $\epsilon<\epsilon_c$ (i.e. at sufficiently low
energies) there are not even local entropy extrema.
\end{enumerate}
The standard interpretation \cite{lyndenbell68} is that, while at
high energies an equilibrium configuration is achievable in the
form of an isothermal sphere, below $\epsilon_c$ the system
collapses to a dense core and overheats, following the
negative-specific-heat trend to its extreme consequences. Such a
``transition'', occurring at the Antonov \emph{point}
$\epsilon=\epsilon_c$, is known as ``gravothermal collapse'' or
``catastrophe'' (see also \cite{lblb,katz} for a more formal
treatment of this transition).

In this work we extend the above theory to rotating systems, by
including the effects of the angular momentum
$\boldsymbol{L}=\sum_{i=1}^N\boldsymbol{r}_i\times\boldsymbol{p}_i$
as a second conserved quantity besides $E$, and discuss the
thermodynamics and the corresponding equilibrium states (which
will now depend on both $E$ and $\boldsymbol{L}$). Our work is
motivated mainly by the crucial role rotation is expected to play
in astrophysical systems on different scales such as globular
clusters and the cores of elliptic galaxies
\cite{chandrasekhar39,binney&tremaine87}, both of which can be
modeled by (\ref{ham},\ref{pot}). (Hence our ``particles'' can be
interpreted equally well as atoms and as stars.) A gravity-driven
collapse can be seen as the prime cause for the equilibrium
shaping of self-gravitating systems, both rotating and not. But
the nature and the result of this collapse are expected to change
significantly in the presence of rotation, and in particular for
systems rotating sufficiently fast. Rotating systems were tackled
in the past by different models (see e.g. \cite{hachisu1,hachisu2}
for a discussion of the stability of rotating gaseous cylinders,
and the more recent works \cite{gross181,slowlyrot} and references
therein), but a complete analysis of the Antonov problem with
angular momentum is still lacking. Some progress was achieved
recently for another class of systems where particles are assumed
to obey a statistics that prohibits overlapping, the so-called
Lynden-Bell statistics \cite{prl,epjb}.

Indeed, we will show that the above picture (a)-(e) holds with
minor modifications for slowly rotating systems, that is when
$|\boldsymbol{L}|=L$ is sufficiently small. In particular,
defining the reduced angular momentum $\lambda=L/\sqrt{RGN^3}$, we
identify an Antonov \emph{line} $\epsilon_c(\lambda)$, with
$\epsilon_c(0)=\epsilon_c\simeq -0.335$ which plays in rotating
systems the same role as the Antonov point plays in non-rotating
ones. For $\lambda$ larger than a critical value $\lambda_c\simeq
0.45$, however, the collapse occurs not to a single dense cluster
but to a double cluster (a ``binary star''), signaling a breaking
of the (axial-)rotational symmetry of (\ref{ham}). It is
remarkable that, strictly speaking, for $\lambda>\lambda_c$ there
is no Antonov limit: double cluster solutions exist at all
energies, albeit being locally unstable (they are saddle points of
entropy in the space of $\rho$'s). The lower the energy, the more
dense and point-like the two clusters. We calculate the global
phase diagram of the model in the $(\epsilon,\lambda)$ plane and
analyze the thermodynamics by studying the phase transition that
take place and the caloric curves. The equilibrium density
profiles obtained in the different regions of the phase diagram
show that, depending on the values of $\epsilon$ and $\lambda$,
the formation of ``rings'', ``single stars'', ``double stars'' and
``disk''-like structures is possible.

From a broad statistical mechanics viewpoint, it is well known
that self-gravitating systems, and more generally systems with
long-range interaction potentials\footnote{By which we mean
potentials decaying with the interparticle distance $r$ more
slowly than $r^{-D-\delta}$ with $\delta>0$ in $D$ dimensions when
$r\to\infty$.}, represent a subtle and delicate issue
\cite{gross174}. For a recent critical discussion of the general
statistical mechanics of non-trivial systems see
\cite{leshouches}. The main problem is that the usual
thermodynamic limit $(N\to\infty,V\to\infty,N/V\text{ finite})$
for systems with potentials such as (\ref{pot}) does not exist
\cite{gallavotti99}. Their thermodynamics is evidently not
extensive. Now extensive systems necessarily have positive
specific heat, but this need not be true for non-extensive ones
\cite{thirring70,gross82,gross158,gross124,gross140,lyndenbell99}.
Hence the canonical and microcanonical description of
non-extensive systems do not necessarily coincide (they do only in
``regular'' phases with positive specific heat, and in the dilute
limit $(N\to\infty,V\to\infty,N/V^{1/3}$ finite$)$ discussed at
length in \cite{devega02a,devega02b}). In the most general and
physically interesting cases, self-gravitating systems should be
studied in the microcanonical ensemble, a circumstance that
obviously introduces several technical difficulties which are
amplified by the introduction of rotation.

The choice of Boltzmann statistics, which is made by most authors
\cite{padmanabhan90}, to describe the behaviour of particles
contributes to complicate the situation further. The basic
assumption is that all particles can eventually occupy the same
point/cell in space, i.e. overlapping between particles is
admitted. With Boltzmann statistics, \emph{the system
(\ref{ham},\ref{pot}) has no ground state}, hence the entropy is
not bounded. This is the ultimate origin of Antonov's catastrophe,
even (as we shall see) in rotating systems. Clearly, if one wants
to describe a system where Newtonian gravity is the dominant
macroscopic interaction then configurations with high particle
density should be avoided, for at such short distances quantum
effects and nuclear interactions become more important than
gravity. One way around this problem is to introduce hard cores
for particles. Another substantially equivalent way is to use
Lynden-Bell statistics \cite{lyndenbell67} instead of Boltzmann's.

We shall proceed as follows. In Sec. 2, we will expound our
microcanonical mean-field theory, deriving the analogous of Eq.
(\ref{is}) for rotating systems. In Sec. 3 we will show results
from the numerical solution of the new equation, and discuss in
particular the phase diagram in the $(\epsilon,\lambda)$ plane and
the caloric curves. A comparison between the Lynden-Bell scenario
discussed in \cite{prl,epjb} and the Boltzmann theory is presented
in Sec. 4. Finally, we formulate our conclusions with some final
remarks .

\section{Mean-field theory}

The theoretical analysis of Antonov's problem in a rotating
framework can be carried out along the lines of the mean-field
approximation described in \cite{prl,epjb}, where the equilibrium
properties of a self-gravitating and rotating system were studied
assuming Lynden-Bell statistics for particles\footnote{A
calculation similar to the one presented here (with Boltzmann
statistics) was performed in \cite{laliena98}, where however the
resulting stationarity condition is studied in the context of the
Thirring model \cite{thirring70}, a simplified model of a rotating
self-gravitating gas.}. The microcanonical entropy in our case is
given by
\begin{equation}
S_N\equiv S_N(E,\boldsymbol{L})=\log\l[\frac{\alpha}{N!}\int
\delta(H_N-E)~
\delta(\boldsymbol{L}-\sum_{i=1}^N\boldsymbol{r}_i\times\boldsymbol{p}_i)
~D\boldsymbol{r}~D\boldsymbol{p}\r]
\end{equation}
where $\alpha$ is a constant that makes the argument of the
logarithm dimensionless. The integral over momenta, namely
\begin{equation}
F_N(\{\boldsymbol{r}_i\},K,\boldsymbol{L})=\int
\delta(K-\frac{1}{2}\sum_{i=1}^N
p_i^2)~\delta(\boldsymbol{L}-\sum_{i=1}^N\boldsymbol{r}_i
\times\boldsymbol{p}_i)~D\boldsymbol{p}
\end{equation}
can be calculated via its Laplace transform (integrals resulting
from insertion of the integral representation of the
$\delta$-function are at most of Gaussian type, hence trivial
\cite{prl,epjb,laliena98})
\begin{eqnarray}
\widetilde{F}_N(\{\boldsymbol{r}_i\},s,\boldsymbol{L}) & = &
\int_0^\infty F_N(\{\boldsymbol{r}_i\},K,\boldsymbol{L})
~e^{-sK}~dK\qquad(\Re s>0)\\
 & = & (\det\mathbb{I})^{-1/2}\l(\frac{2\pi}{s}\r)^{\frac{3N-3}{2}}e^{-\frac{1}{2}s\boldsymbol{L}^T
\mathbb{I}^{-1}\boldsymbol{L}}\label{lapla}
\end{eqnarray}
where $\mathbb{I}\equiv\mathbb{I}(\{\boldsymbol{r}_i\})$ denotes
the inertia tensor, with elements ($a,b=1,2,3$)
\begin{equation}\label{inert}
I_{ab}(\{\boldsymbol{r}_i\})=\sum_{i=1}^N
(r_i^2\delta_{ab}-r_{i,a} r_{i,b})
\end{equation}
and $\boldsymbol{L}^T \mathbb{I}^{-1}\boldsymbol{L}=\sum_{a,b}L_a
I^{-1}_{ab}L_b$. Inversion of (\ref{lapla}) gives
\begin{equation}
F_N(\{\boldsymbol{r}_i\},K,\boldsymbol{L})=
\begin{cases}
\frac{(2\pi)^{\frac{3N-3}{2}}}{\Gamma(\frac{3N-3}{2})\sqrt{\det\mathbb{I}}}(K-\frac{1}{2}\boldsymbol{L}^T
\mathbb{I}^{-1}\boldsymbol{L})^{\frac{3N-5}{2}}
&\text{$K>\frac{1}{2}\boldsymbol{L}^T
\mathbb{I}^{-1}\boldsymbol{L}$}\\ 0&\text{otherwise}
\end{cases}
\end{equation}
hence one is left with
\begin{equation}\label{midd}
S_N=\log\l[\frac{\alpha
A}{N!}\int\frac{[E-\frac{1}{2}\boldsymbol{L}^T
\mathbb{I}^{-1}\boldsymbol{L}-\Phi(\{\boldsymbol{r}_i\})]^{\frac{3N-5}{2}}
}{\sqrt{\det\mathbb{I}}} D\boldsymbol{r}\r]
\end{equation}
where $A=(2\pi)^{\frac{3N-3}{2}}/\Gamma((3N-3)/2)$. If one is
interested (as we are) in the large $N$ limit, one can keep only
the terms of dominating order in $N$ in (\ref{midd}). This yields
\begin{equation}\label{inta}
S_N=\log\l[\frac{\alpha A}{N!}\int[E-\frac{1}{2}\boldsymbol{L}^T
\mathbb{I}^{-1}\boldsymbol{L}-\Phi(\{\boldsymbol{r}_i\})]^{\frac{3N}{2}}
D\boldsymbol{r}\r]
\end{equation}
and it remains to integrate over $V^N$.

To this aim, we partition the spherical box $V$ into $K$ identical
cells labeled by the positions $\boldsymbol{r}_k$ of their centers
($k=1,\ldots,K$), and we let $n(\boldsymbol{r}_k)$ denote the
number of particles inside the cell at position
$\boldsymbol{r}_k$, with $\sum_{k=1}^K n(\boldsymbol{r}_k)=N$. In
order to turn the integral into a sum over the cells, we adopt the
following mean-field approximation for the potential $\Phi$ and
the components of the inertia tensor as functionals of the density
profile $\rho\equiv \rho(\boldsymbol{r})=(K/V)n(\boldsymbol{r})$:
\begin{gather}
\Phi(\{\boldsymbol{r}_i\})~\to~
\Phi[\rho]=-\frac{G}{2}\int\frac{\rho(\boldsymbol{r})\rho(\boldsymbol{r'})}{
|\boldsymbol{r}-\boldsymbol{r'}|}d\boldsymbol{r}d\boldsymbol{r'}\label{potmf}\\
I_{ab}(\{\boldsymbol{r}_i\})~\to~ I_{ab}[\rho]=\int
\rho(\boldsymbol{r})\l(r^2\delta_{ab}-r_a r_b\r)d\boldsymbol{r}
\end{gather}
This choice automatically excludes correlations from the theory.
The microcanonical mean-field entropy can now be written as
\begin{equation}\label{wu3}
S_N^{\text{mf}}=\log\l[\frac{\alpha A
}{N!}\int[E-\frac{1}{2}\boldsymbol{L}^T
\mathbb{I}^{-1}\boldsymbol{L}-\Phi[\rho]]^{\frac{3N}{2}}
P[\rho]~d\rho(\boldsymbol{r})\r]
\end{equation}
where $P[\rho]$ is the probability to observe a density profile
$\rho$. Now $P[\rho]$ is just proportional to the number of ways
in which our $N$ particles can be distributed into the $K$ cells,
and is thus determined by the statistical model by which one
describes the behaviour of the particles. Assuming Boltzmann
statistics, there is no spatial constraint for particles so that
all of them can eventually occupy the same cell in $V$. This leads
to
\begin{equation}\label{statis}
P[\rho]\propto\frac{N!}{n(\boldsymbol{r}_1)!\cdots
n(\boldsymbol{r}_K)}\simeq N!~\exp\l[-\frac{NK}{V}\int
c(\boldsymbol{r})(\log c(\boldsymbol{r})-1)d\boldsymbol{r}\r]
\end{equation}
where in obtaining the last expression we used Stirling's
approximation and defined the relative cell occupancy
$c(\boldsymbol{r})=n(\boldsymbol{r})/N=V\rho(\boldsymbol{r})/(KN)$.
Introducing the dimensionless variable
$\boldsymbol{x}=\boldsymbol{r}/R$ and neglecting irrelevant
constants, one finally arrives at
\begin{gather}
S_N^{\text{mf}}=\log\int e^{N\Sigma[c]}~dc(\boldsymbol{x})\\
\Sigma[c]=\frac{3}{2}\log\l[E-\frac{1}{2}\boldsymbol{L}^T
\mathbb{I}^{-1}\boldsymbol{L}-\Phi[c]\r]-\frac{1}{\Theta}\int
c(\boldsymbol{x})(\log c(\boldsymbol{x})-1)d\boldsymbol{x}
\end{gather}
where $\Theta=\int
c(\boldsymbol{x})d\boldsymbol{x}=\frac{V}{KR^3}$.

For $N$ large though finite, $S_N^{\text{mf}}$ can be estimated by
the steepest descent method, which for $S=S_N^{\text{mf}}/N$
yields
\begin{equation}
S=\max_{c(\boldsymbol{x})}\l[\frac{3}{2}\log\l[E-\frac{1}{2}\boldsymbol{L}^T
\mathbb{I}^{-1}\boldsymbol{L}-\Phi[c]\r]-\frac{1}{\Theta}\int
c(\boldsymbol{x})(\log c(\boldsymbol{x})-1) d\boldsymbol{x}\r]
\end{equation}
Notice that the stationarity properties of $S_N$ do not depend on
$\Theta$. This is most easily seen by setting
$f(\boldsymbol{x})=c(\boldsymbol{x})/\Theta$, which leads
to\begin{equation}
S=\max_{f(\boldsymbol{x})}\l[\frac{3}{2}\log\l[E-\frac{1}{2}\boldsymbol{L}^T
\mathbb{I}^{-1}\boldsymbol{L}-\Phi[f]\r]-\int
f(\boldsymbol{x})(\log f(\boldsymbol{x})-1)
d\boldsymbol{x}+\log\Theta\r]
\end{equation}
We remark that $\mathbb{I}\equiv\mathbb{I}[f]$. In terms of $f$
and in units of $GN^2/R$ and $NR^2$, respectively, $\Phi$ and
$I_{ab}$ (see (\ref{potmf}),(\ref{inert})) become simply
\begin{gather}
\Phi[f]=-\frac{1}{2}\int\frac{f(\boldsymbol{x})f(\boldsymbol{x'})
}{|\boldsymbol{x}-\boldsymbol{x'}|}d\boldsymbol{x}d\boldsymbol{x'}\\
I_{ab}[f]=\int f(\boldsymbol{x})(x^2\delta_{ab}-x_a
x_b)d\boldsymbol{x}
\end{gather}
The stationarity condition $\delta S/\delta f=0$ implies the
following equation for the extrema of the entropy:
\begin{equation}\label{equa}
f(\boldsymbol{x})=\exp\l[\beta\int\frac{f(\boldsymbol{x'})}{
|\boldsymbol{x}-\boldsymbol{x'}|}~d\boldsymbol{x'}+\frac{\beta}{2}
(\boldsymbol{\omega}\times\boldsymbol{x})^2+\mu\r]
\end{equation}
where $\boldsymbol{\omega}=\mathbb{I}^{-1}\boldsymbol{L}$ is the
angular velocity and $\mu$ is a Lagrange multiplier that ensures
the constraint $\int f(\boldsymbol{x})d\boldsymbol{x}=1$, and
\begin{equation}
\beta=\frac{3/2}{[E-\frac{1}{2}\boldsymbol{L}^T
\mathbb{I}^{-1}\boldsymbol{L}-\Phi[f]]}
\end{equation}
is the inverse kinetic energy. Notice that $\beta\equiv\beta[f]$.

Eq. (\ref{equa}) is the analogous of (\ref{is}) for a rotating
system. Unfortunately, it cannot be solved analytically. However,
some simple qualitative considerations are possible. In absence of
rotation, the solution of (\ref{equa}) is known to depend only on
$x=|\boldsymbol{x}|$. The minimum expected effect of rotation is
to create distortions that cannot be described by $x$ alone. Hence
it is necessary to introduce angular variables, and it is
convenient to pass to spherical coordinates $(x,\theta,\phi)$. If
the system is slowly rotating and if, as we shall assume in the
following, the angular momentum is parallel to the $3$-axis, one
may neglect the $\phi$-dependence and write formally the solution
of (\ref{equa}) as a sum of Legendre polynomials weighted by
functions of $x$ alone\footnote{When in the following we speak of
axial rotational symmetry or for short of axial symmetry we refer
always to the $3$-axis.}:
\begin{equation}
f(\boldsymbol{x})=\sum_{l=0}^\infty f_l(x) P_l(\cos\theta)
\end{equation}
This procedure is well-known \cite{chandrasekhar33}, and has been
recently applied to the Antonov problem for slowly rotating
systems by Chavanis \cite{slowlyrot}. The latter, however, do not
display any serious qualitative difference from non rotating ones.
The most interesting phenomena occur at high angular momenta, and
in order to describe them it is necessary to include the
dependence on $\phi$. We thus write the solution of (\ref{equa})
as
\begin{equation}
f(\boldsymbol{x})=\sum_{l=0}^\infty \sum_{m=-l}^l
f_{l,m}(x)Y_{lm}(\theta,\phi)\label{ser}
\end{equation}
where $Y_{lm}$ are real spherical harmonics and $f_{l,m}$ are
their weights (again functions of $x$ alone), which will have to
be determined. Clearly, this expansion differs from the one in
Legendre polynomials essentially for the terms with $m\neq 0$. We
shall see that these terms are the ones that provide the relevant
new physics at high angular momenta.

Using this expansion it is easy to show the following statements.
The normalization $\int f(\boldsymbol{x})d\boldsymbol{x}=1$ is
equivalent to
\begin{equation}
2\sqrt{\pi}\int x^2 f_{0,0}(x) dx=1
\end{equation}
so that the $(0,0)$ harmonics is ensures this condition. The
coordinates of the center of mass are given by
\begin{gather}
x_1^{CM}\equiv\int x_1
f(\boldsymbol{x})d\boldsymbol{x}=2\sqrt{\frac{\pi}{3}}\int x^3
f_{1,1}(x)dx\nonumber\\ x_2^{CM}\equiv\int x_2
f(\boldsymbol{x})d\boldsymbol{x}=2\sqrt{\frac{\pi}{3}}\int x^3
f_{1,-1}(x)dx\label{cm}\\ x_3^{CM}\equiv\int x_3
f(\boldsymbol{x})d\boldsymbol{x}=2\sqrt{\frac{\pi}{3}}\int x^3
f_{1,0}(x)dx\nonumber
\end{gather}
so that harmonics with $l=1$ are responsible for fixing the center
of mass. It is convenient to set the latter in the center of the
spherical box ($\boldsymbol{x}=\boldsymbol{0}$), which amounts to
postulating $f_{1,m}=0$ for $m\in\{-1,0,1\}$. The diagonal
components of the inertia matrix read\footnote{We warn that
indices like in $I_{ab}$ refer to spatial components, while
indices like $f_{l,m}$ refer to harmonics.}
\begin{gather}
I_{11}=\sqrt{\frac{16\pi}{9}}\int x^2
f_{0,0}(x)dx+\sqrt{\frac{4\pi}{45}}\int x^4 f_{2,0}(x)
dx-\sqrt{\frac{4\pi}{15}}\int x^4 f_{2,2}(x) dx\\
I_{22}=\sqrt{\frac{16\pi}{9}}\int x^2
f_{0,0}(x)dx+\sqrt{\frac{4\pi}{45}}\int x^4 f_{2,0}(x)
dx+\sqrt{\frac{4\pi}{15}}\int x^4 f_{2,2}(x) dx\\
I_{33}=\sqrt{\frac{16\pi}{9}}\int x^2
f_{0,0}(x)dx-\sqrt{\frac{16\pi}{45}}\int x^4 f_{2,0}(x) dx
\end{gather}
while for the off-diagonal elements one gets
\begin{gather}
I_{12}=-\sqrt{\frac{4\pi}{15}}\int x^4 f_{2,-2}(x) dx\\
I_{13}=-\sqrt{\frac{4\pi}{15}}\int x^4 f_{2,1}(x) dx\\
I_{23}=-\sqrt{\frac{4\pi}{15}}\int x^4 f_{2,-1}(x) dx
\end{gather}
In particular, one has
\begin{equation}\label{armopar}
I_{11}-I_{22}=-\sqrt{\frac{16\pi}{15}}\int x^4 f_{2,2}(x)dx
\end{equation}
so that this parameter, which will play a foremost role in the
following analysis, only depends on the $(2,2)$ harmonics.

Using spherical harmonics as a basis set, for the Newtonian
potential one has \cite{jackson75,wyld76}:
\begin{equation}\label{newta}
\frac{1}{|\boldsymbol{x}-\boldsymbol{x'}|}\!=\!\sum_{l=0}^\infty
\!\sum_{m=-l}^l\!\frac{4\pi}{2l+1}\frac{\min\{x,x'\}^l}{\max\{x,x'\}^{l+1}}
Y_{lm}(\theta,\phi)Y_{lm}(\theta',\phi')
\end{equation}
Now it is easy to see that the so far unknown functions $f_{l,m}$
satisfy the following integral equation (which can be found by
multiplying both sides of (\ref{equa}) by $Y_{lm}$ and integrating
over angular variables):
\begin{equation}\label{arra}
f_{l,m}(x)=\int Y_{lm}(\theta,\phi)~e^{\beta \sum_{l',m'}
u_{l',m'}(x)Y_{l'm'}(\theta,\phi) +\frac{1}{2}\beta\omega^2 x^2
\sin^2\theta} d\theta d\phi
\end{equation}
with
\begin{equation}\label{uh}
u_{l,m}(x)=\frac{4\pi}{2l+1}\int\frac{\min\{x,x'\}^l}{\max\{x,x'\}^{l+1}}
~f_{l,m}(x')(x')^2 dx'
\end{equation}
Despite their unfriendly look, Eqs (\ref{arra},\ref{uh}), which
should be solved together in a self-consistent way, can be dealt
with numerically. Technical details are similar to those described
in \cite{prl,epjb} in relation to the Lynden-Bell statistics case,
and we will skip them here. Let it suffice to say that for the
computation the series (\ref{newta}) was truncated at the order
$l_{\text{max}}=12$ and that all harmonics (even and odd) up to
$l_{\text{max}}$ were included, except those with $l=1$. Notice
that this does not imply a truncation of the expansion
(\ref{ser}), i.e. the $f_{l,m}$ exist also for orders higher than
$l_{\text{max}}$. The effects of truncating the potential are
well-known, and are discussed e.g. in \cite{epjb}. In short, when
truncated at higher and higher order the series (\ref{newta})
differs more and more from the real potential in a narrower and
narrower region around $\boldsymbol{x}=\boldsymbol{x'}$, in a way
that is reminiscent of the Gibbs phenomenon. As we said, the
angular momentum was taken to lie along the $3$-direction. In this
setup, solutions of (\ref{equa}) were calculated as functions of
$\epsilon=\frac{ER}{GN^2}$ and $\lambda=\frac{L}{\sqrt{RGN^3}}$;
results of this analysis are given and discussed in the next
section.

\section{Results}

\subsection{Relevant quantities}

The crucial quantity in the analysis that will follow is given by
\begin{equation}\label{hess}
{\rm Hes}(\epsilon,\lambda)={\rm det}
\begin{pmatrix} \partial^2_\epsilon S &
\partial_\lambda\partial_\epsilon
S
\\ \partial_\epsilon\partial_\lambda S & \partial_\lambda^2 S
\end{pmatrix}=\kappa_1\kappa_2
\end{equation}
where $\kappa_{1,2}$ are the eigenvalues of the matrix, ordered
assuming $\kappa_1>\kappa_2$. The relevance of ${\rm
Hes}(\epsilon,\lambda)$ lies in the fact that it relates the
thermostatistical equilibrium properties of systems in the
microcanonical ensemble to the topology of the entropy surface in
the space of conserved macroscopic quantities. In general, it can
be shown \cite{gross174,gross186} that pure, stable thermodynamic
phases with positive heat capacity are characterized by having
$\kappa_1<0$ (hence ${\rm Hes}(\epsilon,\lambda)>0$). Here, $S$ as
a function of both $\epsilon$ and $\lambda$ is concave, and
statistical ensembles (microcanonical and canonical) are
equivalent. When at least $\kappa_1>0$, one has a phase
coexistence region with negative specific heat, and statistical
ensembles are inequivalent. $S(\epsilon,\lambda)$ has a convex
part that is ignored by the canonical ensemble, which considers
the concave hull of $S$. The convex region corresponds to negative
specific heat. For self-gravitating systems, (\ref{hess}) provides
a powerful characterization of some of the relevant quantities. In
particular, we shall concentrate on the following aspects of the
phase diagram.

\emph{Antonov limit.} For each fixed $\lambda$, the Antonov limit
$\epsilon_c(\lambda)$ is the minimum $\epsilon$ for which a
solution of (\ref{equa}) exists. In general, $\epsilon_c(\lambda)$
can be \emph{defined} by the condition
\begin{equation}\label{al}
\lim_{\epsilon\downarrow\epsilon_c(\lambda)}\kappa_1=+\infty
\end{equation}
(Strictly speaking, $\kappa_1$ can diverge only in the mean-field
approximation in the $N\to\infty$ limit at fixed $\epsilon$ and
$\lambda$.) This is a straightforward generalization of the
non-rotating case, where at the Antonov point
$\epsilon_c\equiv\epsilon_c(0)$ one has
\begin{equation}
\lim_{\epsilon\downarrow\epsilon_c}\frac{\partial\beta}{\partial\epsilon}=+\infty
\end{equation}
When two (or more) conserved quantities have to be taken into
consideration, the above condition becomes a condition on the
maximum eigenvalue of the matrix of second derivatives of $S$ with
respect to the conserved quantities. Upon varying $\lambda$, the
Antonov limit $\epsilon_c(\lambda)$ describes a line in the
$(\epsilon,\lambda)$ plane. We call this line the Antonov line.

\emph{Phase coexistence.} By virtue of the properties of
(\ref{hess}), starting in the high-energy phase and decreasing
$\epsilon$ at fixed $\lambda$ one finds the beginning of a phase
coexistence region when
\begin{equation}\label{hesz}
{\rm Hes}(\epsilon,\lambda)=0
\end{equation}
is fulfilled. We shall denote by $\epsilon_0(\lambda)$ the
corresponding value of $\epsilon$. In particular, at
$\epsilon_0(\lambda)$ the specific heat at constant
$\beta\omega=-\frac{\partial S}{\partial\lambda}$
($\omega=|\boldsymbol{\omega}|$), i.e.
\begin{equation}\label{cc}
c_{\beta\omega}=-\frac{\beta^2}{{\rm
Hes}(\epsilon,\lambda)}\frac{\partial^2 S}{\partial\lambda^2}
\end{equation}
has a pole and changes sign (from positive to negative) upon
decreasing $\epsilon$. Again, varying $\lambda$ the point
$\epsilon_0(\lambda)$ describes a line in the $(\epsilon,\lambda)$
plane (with $\epsilon_0(0)\simeq -0.2$), which separates a pure
thermodynamic phase from a phase coexistence region.

\emph{Bifurcation to axially asymmetric solutions.} For
sufficiently high $\lambda$, solutions of (\ref{equa}) without
axial rotational symmetry (``binary stars'') bifurcate
(continuously) from axially-symmetric solutions \cite{prl,epjb}.
This bifurcation can be detected with the aim of the order
parameter (see (\ref{armopar}))
\begin{equation}
D=|I_{11}-I_{22}|=\l|\sqrt{\frac{16\pi}{15}}\int x^4
f_{2,2}(x)dx\r|
\end{equation}
introduced in \cite{prl}. It is easy to understand that solutions
with $D=0$ are axially-symmetric, while solutions with $D\neq 0$
break axial rotational symmetry. If $\lambda=0$ the system is
necessarily isotropic ($I_{11}=I_{22}=I_{33}$) and spherical
symmetry cannot be broken. This reflects the fact that the
solution of (\ref{equa}) depends only on $x$, so that the series
expansion (\ref{ser}) is truncated after the first term, which
involves only the $Y_{00}$ harmonics. When $\lambda\neq 0$,
anisotropies may occur ($I_{33}\neq I_{11},I_{22}$) and one can
have either solutions that are rotationally-symmetric around the
$3$-axis ($I_{11}=I_{22}$ or $D=0$), which eventually depend on
$x$ and $\theta$ to account for obvious rotation-induced
distortions, or solutions that are asymmetric around the $3$-axis
($I_{11}\neq I_{22}$ or $D\neq 0$). The latter must depend on $x$,
$\theta$ and $\phi$, and correspond to double clusters. However,
it turns out that such solutions appear only when $\lambda$ is
larger than a threshold value and $\epsilon$ is sufficiently low
(see also the discussion at the end of Sec. 4, and in particular
Fig. \ref{entr}).

\emph{Stability of the double cluster solutions.} Finally, we
shall be interested in the local stability properties of
double-cluster solutions (high-$\epsilon$, axially symmetric
solutions are locally stable because in this limit the kinetic
term of (\ref{ham}) dominates and the system behaves more and more
as an ideal gas). This amounts to looking for the \emph{marginal
stability line} defined by the condition
\begin{equation}\label{localstab}
\l[\frac{\delta ^2 S}{\delta f(\boldsymbol{x})\delta
f(\boldsymbol{x'})}\r]_{G[f^*]=0}=0
\end{equation}
where
\begin{equation}
G[f]=f(\boldsymbol{x})-\exp\l[\beta\int\frac{f(\boldsymbol{x'})}{
|\boldsymbol{x}-\boldsymbol{x'}|}~d\boldsymbol{x'}+\frac{\beta}{2}
(\boldsymbol{\omega}\times\boldsymbol{x})^2+\mu\r]
\end{equation}
and $f^*$ is an axially asymmetric solution of (\ref{equa}). The
marginal stability line marks the border between the region of the
phase diagram where axially asymmetric solution are stable from
that where they are at most metastable (saddle points in the
entropy surface). In principle, at the most general level the
local stability analysis requires solving the eigenvalue problem
for the kernel $K(\boldsymbol{x},\boldsymbol{x'})=\frac{\delta ^2
S}{\delta f(\boldsymbol{x})\delta f(\boldsymbol{x'})}$. We have
not been able to carry out this task. However, it is possible to
get numerical information about (\ref{localstab}). Some technical
details about this problem are given in \cite{epjb}, and we will
not discuss them here.

\begin{figure}
\begin{center}
\includegraphics[width=8cm,angle=-90]{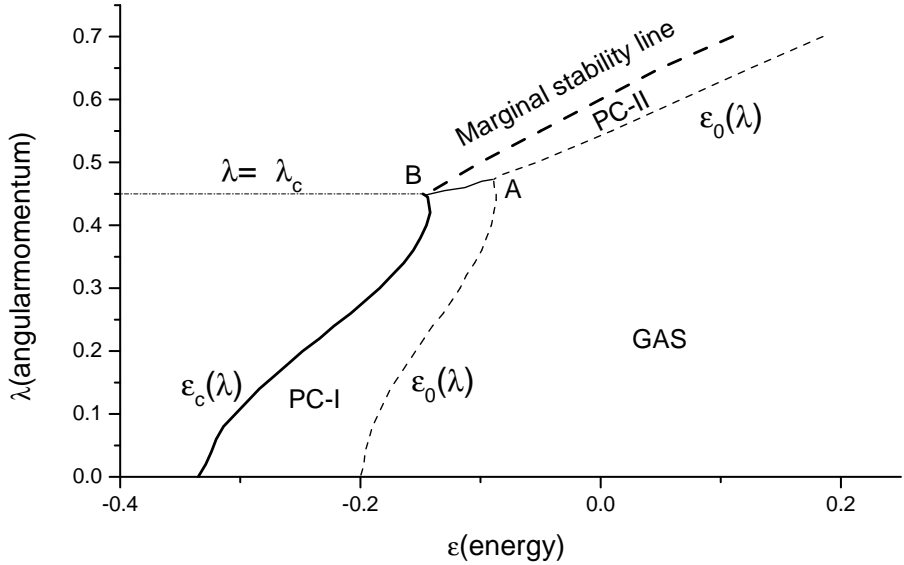}
\caption{\label{pd}Microcanonical phase diagram. GAS (axially
symmetric) denotes the pure phase; phase coexistence regions with
negative specific heat are denoted by PC-I (single-cluster plus
``gas'') and PC-II (double-cluster plus ``gas''). Lines are as
follows. Solid-continuous: Antonov line $\epsilon_c(\lambda)$, Eq.
(\ref{al}) . Light-dashed: phase boundary $\epsilon_o(\lambda)$,
Eq. (\ref{hesz}). The continuous line from A to B marks the
beginning of the bifurcation line for the onset of double cluster
solutions and separates the PC-I and PC-II phases; this line
merges with $\epsilon_0(\lambda)$ right of $A$. Heavy-dashed:
marginal stability line for the double cluster solution. Results
have been obtained with $l_{\text{max}}=12$.}
\end{center}
\end{figure}

\begin{figure}[!hb]
\begin{center}
\subfigure{\scalebox{.53}{\includegraphics{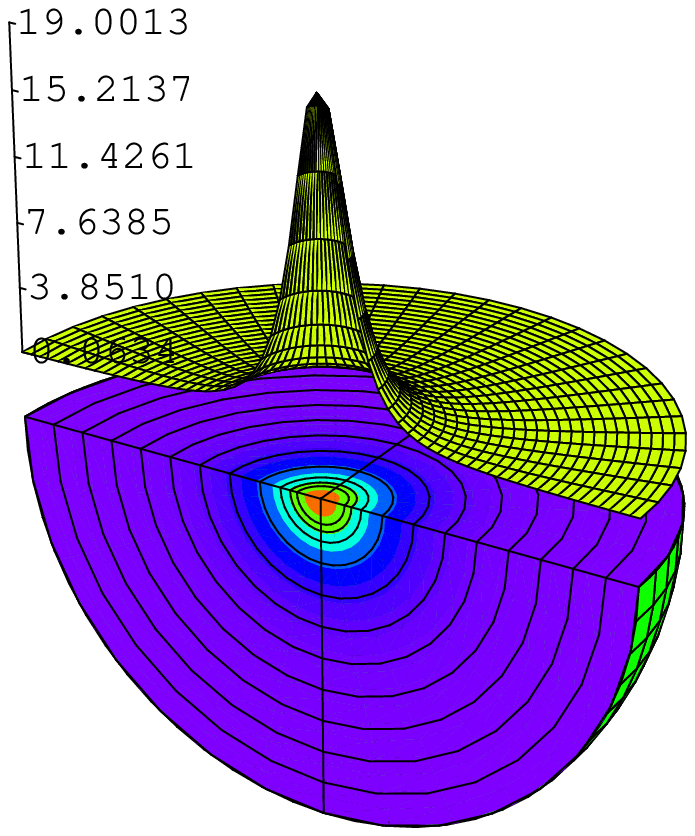}}}
\subfigure{\scalebox{.53}{\includegraphics{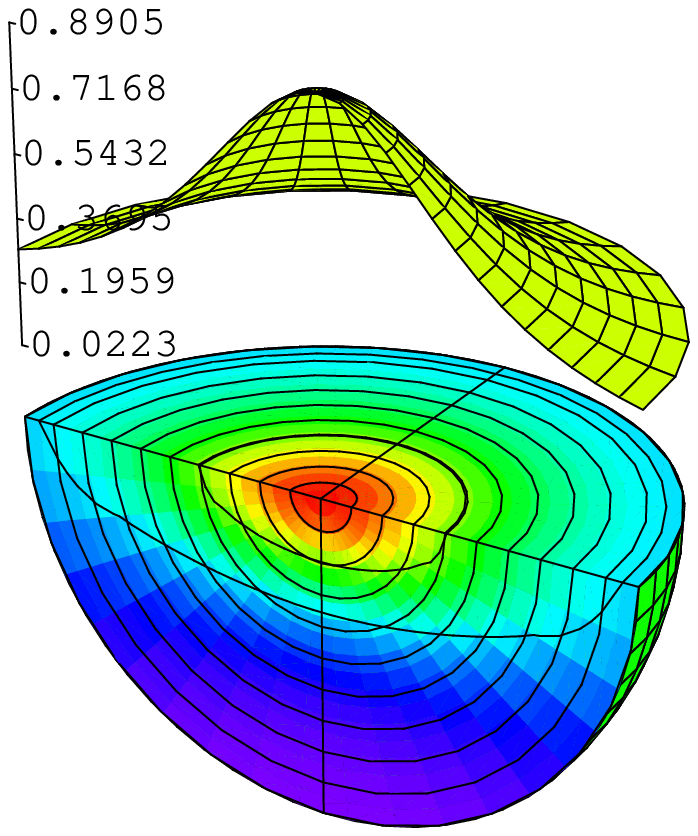}}}
\subfigure{\scalebox{.53}{\includegraphics{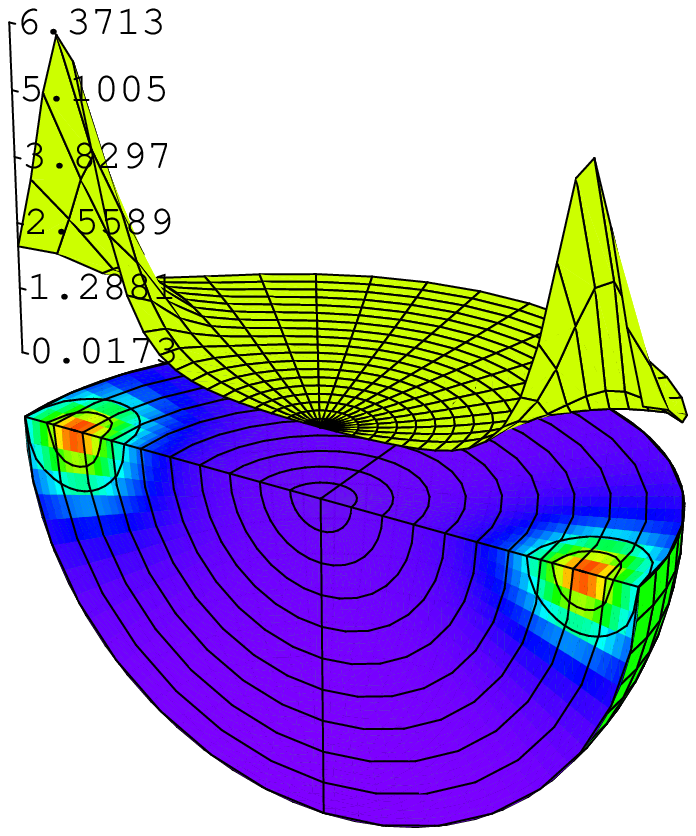}}}
\caption{\label{distri}Example of stable equilibrium
configurations, contour plots and (above) density profiles. Left
to right: single-cluster (collapsed state,
$\epsilon=-0.33,\lambda=0.001$); ``disk''
($\epsilon=-0.1,\lambda=0.4$); double-cluster
($\epsilon=0,\lambda=0.6$).}
\end{center}
\end{figure}

\subsection{Phase diagram}

The phase diagram resulting from numerical solution of
(\ref{equa}) is displayed in Fig. \ref{pd}. In Fig. \ref{distri}
we show instead a small sample of equilibrium density profiles.
The situation for sufficiently low angular $\lambda$ resembles
that of non-rotating systems. At high $\epsilon$, a pure ``gas''
phase is found, where the equilibrium density profiles are
axially-symmetric and particles are uniformly spread over $V$. As
$\epsilon$ is decreased, the Newtonian contribution becomes more
and more important and at the $\epsilon_0(\lambda)$ line, the
specific heat becomes negative and one enters a phase coexistence
region, labeled PC-I, where a collapsed configuration (see Fig.
\ref{distri}a) competes with the gas state (see Fig.
\ref{distri}b). Upon further decreasing $\epsilon$ no more
solutions of (\ref{equa}) are found and the Antonov limit is
reached. In this region, we indeed find that
\begin{equation}
\lim_{\epsilon\downarrow\epsilon_c(\lambda)}{\rm
Hes}(\epsilon,\lambda)=-\infty
\end{equation}
which, comparing with (\ref{al}), implies that $\kappa_2<0$.
Rotation has the only effects of shifting both the onset of
negative heat capacity and the Antonov limit to higher
$\epsilon$'s (as also argued in \cite{slowlyrot}), and of
deforming the density profile (as clear from Fig. \ref{distri}b).

When $\lambda$ becomes larger than the critical value
$\lambda_c\simeq 0.45$, however, the picture changes drastically.
Here axially-symmetric solutions become unstable against
fluctuations that break such a rotational symmetry and
axially-asymmetric solutions, i.e. ``double stars'', occur (see
Fig. \ref{distri}c). The picture is clear for high $\lambda$. Upon
decreasing $\epsilon$, one encounters the $\epsilon_0(\lambda)$
line (which in that region coincides with the bifurcation line)
and enters a phase coexistence region (labeled PC-II) with
negative specific heat $c_{\beta\omega}$, Eq. (\ref{cc})), where
double-cluster and gas configurations compete. However, at odds
with the low-$\lambda$ case, a further decrease in $\epsilon$
makes double-clusters unstable. Remarkably, the solution for lower
$\epsilon$ still exists, hence there is no Antonov limit for
$\lambda>\lambda_c$ strictly speaking. This issue will be further
discussed in the next section.

The bifurcation line and the $\epsilon_0(\lambda)$ line do not
coincide in a small range close to $\lambda_c$ (the AB line in
Fig. \ref{pd} gives the bifurcation line in this range). Here the
bifurcation line separates the two mixed phases, i.e. upon
crossing it by increasing $\lambda$ at fixed $\epsilon$ double
clusters are obtained. Transitions from PC-I to PC-II can also
occur upon decreasing $\epsilon$ at fixed $\lambda$.

It is important to stress that these results were obtained by
truncating the potential via (\ref{newta}) to order
$l_{\text{max}}=12$. While calculations performed with smaller
$l_{\text{max}}$ in the lower half of the phase diagram
($\lambda<\lambda_c$) gave essentially the same results, for high
angular momenta we observed some dependence on $l_{\text{max}}$,
albeit weak. In particular, it is possible that upon increasing
$l_{\text{max}}$ the marginal stability line is further inclined
towards the $\epsilon_0(\lambda)$ line. As a consequence, the
PC-II phase would become thinner and thinner, so that double stars
would be stable (if at all) only in a very tiny portion of the
parameter space. This quantitative issue, which is strictly
related to the use of Boltzmann statistics, certainly deserves
further investigation. We believe that no other main (qualitative)
changes to the picture presented here should occur.

\begin{figure}
\begin{center}
\subfigure[~]{\includegraphics[width=8cm,angle=-90]{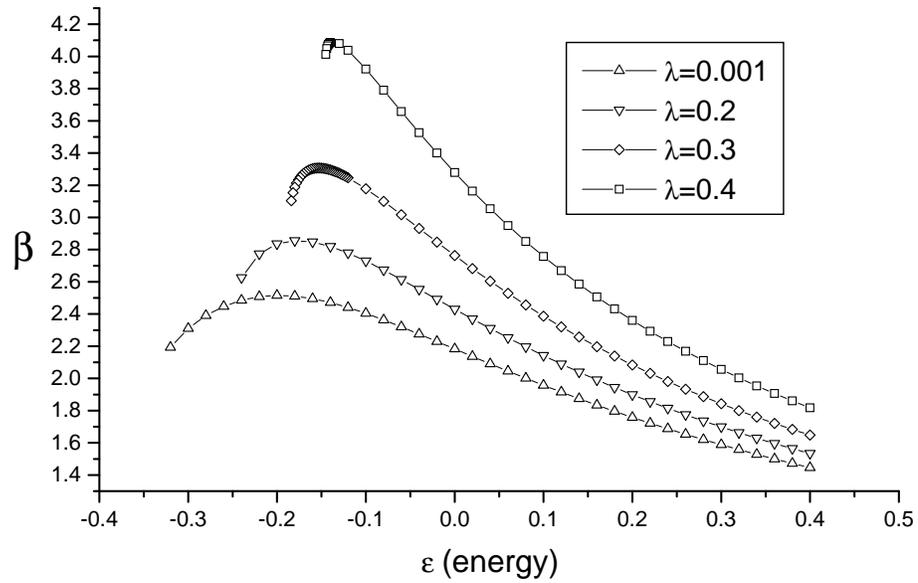}}
\subfigure[~]{\includegraphics[width=8cm,angle=-90]{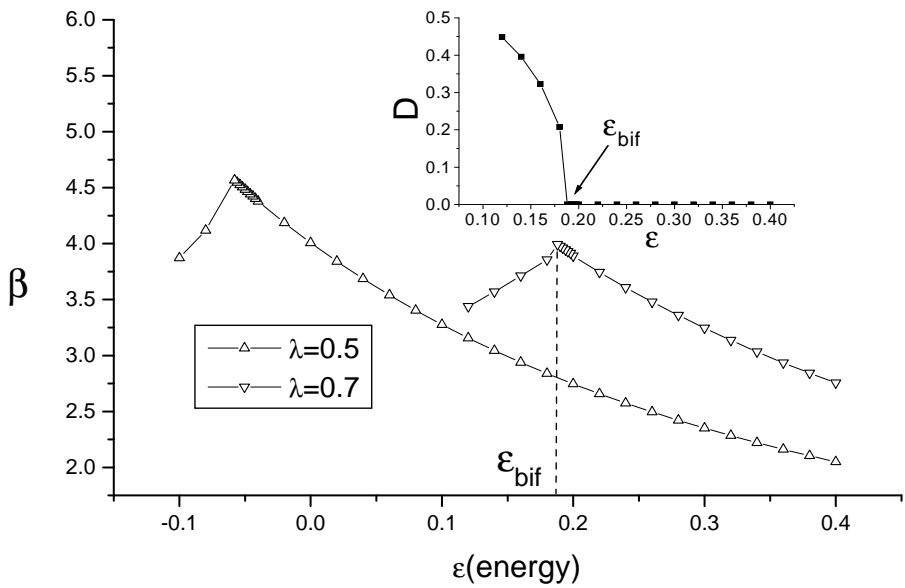}}
\caption{\label{cal}Caloric curves for (a) low and (b) high
$\lambda$. The inset in (b) shows the behaviour of the order
parameter $D$ at $\lambda=0.7$. $\epsilon_{\text{bif}}$ in (b)
denotes the bifurcation energy at $\lambda=0.7$.}
\end{center}
\end{figure}

Finally, moving on to thermodynamics, in Fig. \ref{cal} we show
the caloric curves $\beta$ vs $\epsilon$ at fixed low and,
respectively, high $\lambda$. The peak in (b) corresponds to the
bifurcation line.

\section{Comparison with Lynden-Bell statistics}

It is interesting to compare the behaviour of the self-gravitating
and rotating gas of Boltzmann particles described above with that
of a same system of Lynden Bell particles \cite{prl,epjb}. Using
Lynden-Bell statistics \cite{lyndenbell67} instead of Boltzmann's
amounts to postulating a maximum capacity of $n_0\ll N$ particles
for each of the $K$ cells in which $V$ is subdivided in order to
compute the integral (\ref{inta}), so that Eq. (\ref{statis})
would become
\begin{displaymath} P[\rho]\propto\frac{N!}{n(\boldsymbol{r}_1)!\cdots
n(\boldsymbol{r}_K)!}\prod_{{\rm cells~}k}
\frac{n_0!}{(n_0-n(\boldsymbol{r}_k))!}
\end{displaymath}
From the physical viewpoint, Lynden-Bell systems are leptodermous.
This is not possible for Boltzmann particles, as seen by the fact
that the clusters in Fig. \ref{distri} are sharply peaked, with
the density decreasing smoothly from the maximum (Lynden-Bell
particles can form stars with a dense, finite core and a
fast-decreasing density \cite{prl}). Correspondingly, the
stationary point condition for $c=\Theta f$ reads
\begin{displaymath}
c(\boldsymbol{x})=\l[1+\exp\l[-\frac{\beta}{\Theta}
\int\frac{c(\boldsymbol{x'})}{
|\boldsymbol{x}-\boldsymbol{x'}|}~d\boldsymbol{x'}
-\frac{1}{2}\beta(
\boldsymbol{\omega}\times\boldsymbol{x})^2+\mu\r]\r]^{-1}
\end{displaymath}
This case has been studied thoroughly in \cite{prl,epjb}. Notice
that, at odds with what happens in Boltzmann statistics (which is
the $n_0\to N$ followed by $N\to\infty$ limit of the Lynden-Bell
case), here the result depends on $\Theta$, which is now given by
$\Theta=\frac{NV}{n_0 K R^3}$, and is related to the total mass of
the system.

\begin{figure}
\begin{center}
\subfigure[~]{\includegraphics[width=8cm,angle=-90]{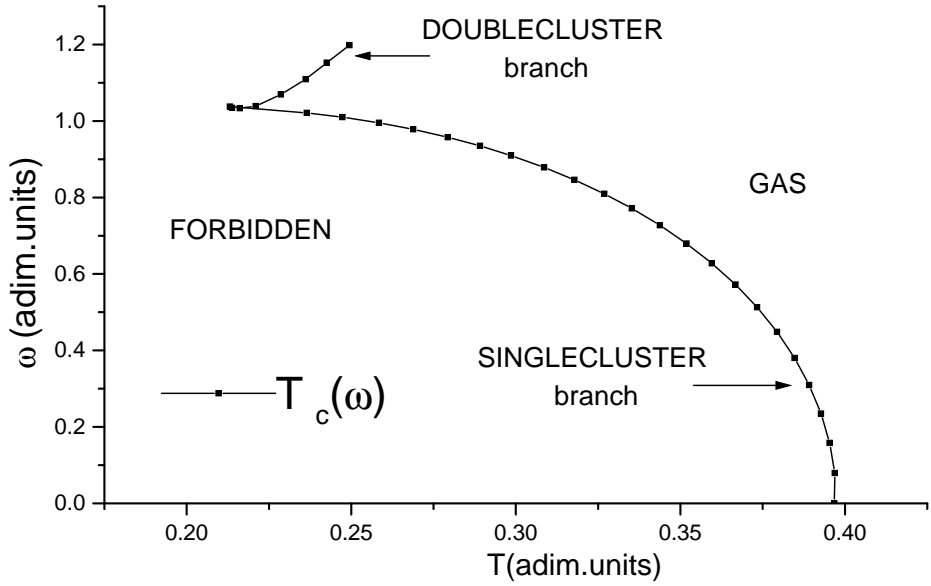}}
\subfigure[~]{\includegraphics[width=8cm,angle=-90]{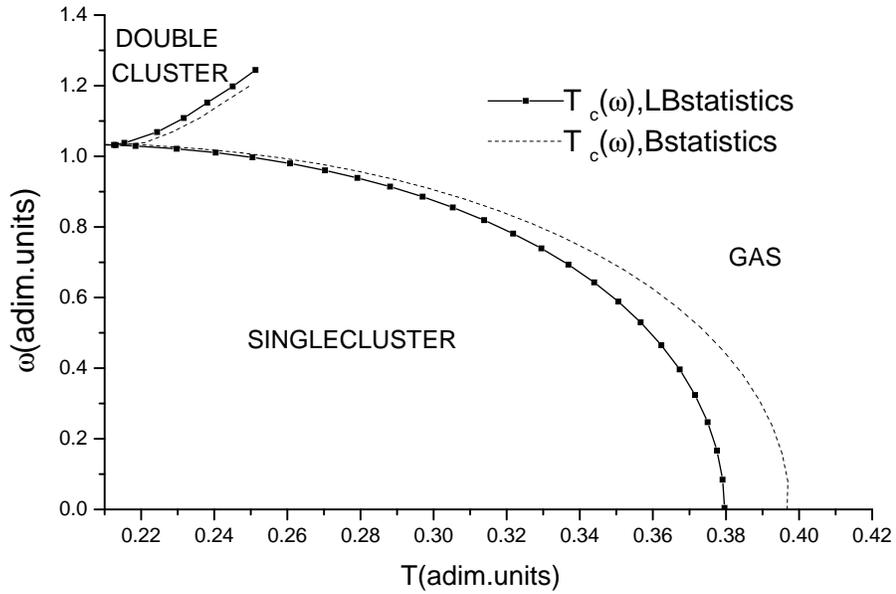}}
\caption{\label{can}Canonical phase diagram for (a) Boltzmann
statistics and (b) Lynden-Bell statistics. The dashed line in (b)
reproduces (a) and shows the difference between the two types of
statistics.}
\end{center}
\end{figure}

The microcanonical phase diagram for this case is discussed in
\cite{prl,epjb}, where it is shown that the system has three pure
phases (''gas'', single cluster and double-cluster) separated by a
large phase coexistence region with negative specific heat. For
historical reasons, in this paper we shall discuss the canonical
phase diagram, which can be obtained by mapping the
$(\epsilon,\lambda)$ variables onto their conjugate intensive
variables $(T\equiv 1/\beta,\omega)$. Evidently, in the canonical
ensemble phase coexistence regions cannot be detected
\cite{gross174}, as one must always have positive specific heat.
This means that the collapse temperature is determined by the
value of $\beta$ at which the derivative of $\beta$ versus
$\epsilon$ changes sign, and the region PC-I, corresponding to
lower energies, is not accessible. For the present comparison
purposes, however, this limited information is sufficient.
Canonical phase diagrams are shown in Fig.~\ref{can}.

In the Boltzmann case, one sees that for high temperatures the
equilibrium state is an axially-symmetric, gas-like one. At a
critical temperature $T_c(\omega)$ the system collapses into a
single cluster (for $\omega<\omega_c\simeq 1.03$) or into a double
cluster (for $\omega>\omega_c$) and lower temperatures are
forbidden. (For $\omega=0$ the known value $T_c(0)\simeq 0.3967$
is recovered.) Lynden-Bell statistics turns out to shift the onset
of both the collapse to a single cluster and the formation of the
double cluster to lower temperatures. However, at lower
temperatures the system can exist in a single-cluster state (low
$\omega$) or in a double cluster state (high $\omega$). Notice how
in both cases the phase coexistence regions with negative specific
heat that are found in the microcanonical phase diagrams are lost
in the canonical setup.

\begin{figure}
\begin{center}
\includegraphics[width=8cm,angle=-90]{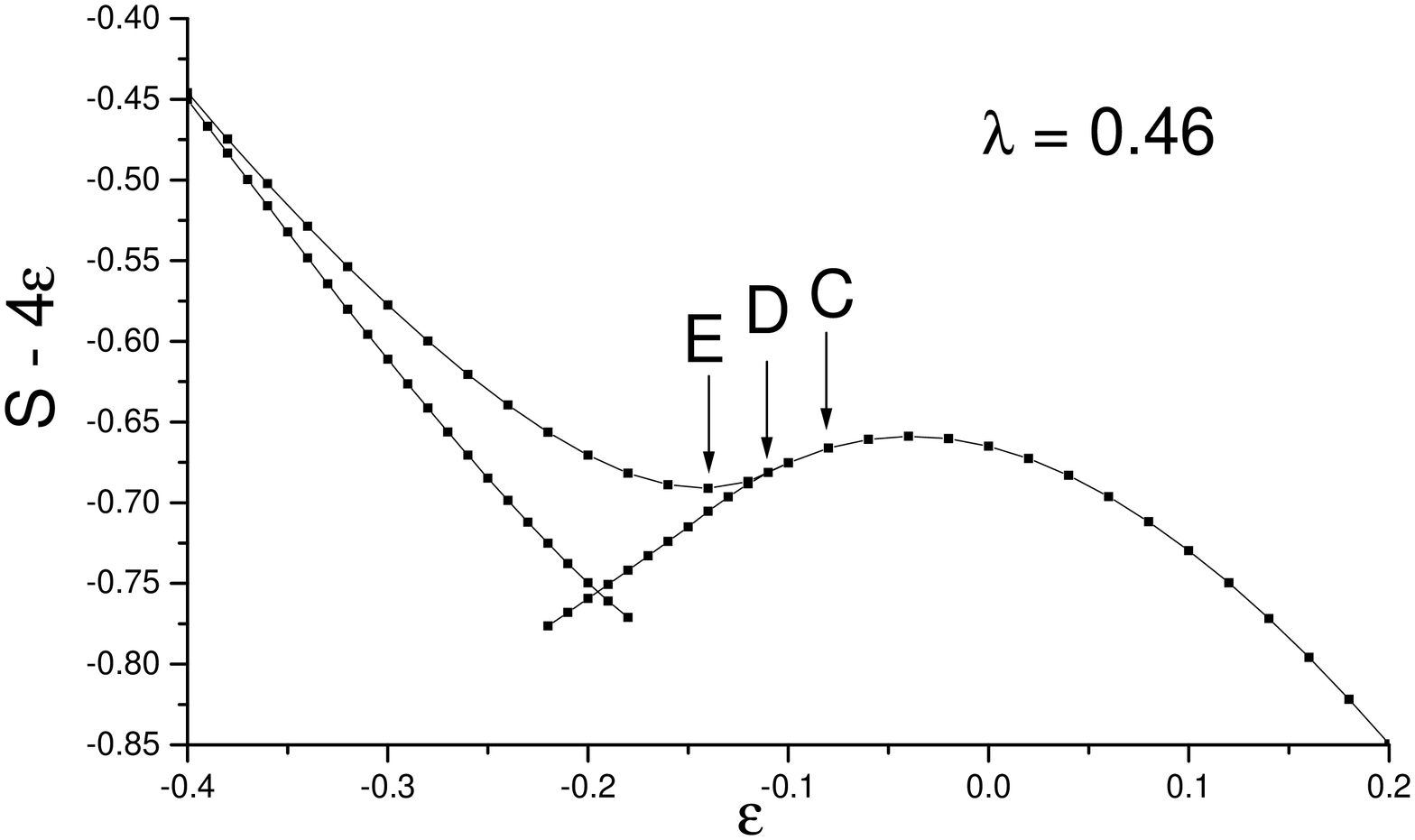}
\caption{\label{entr}(Entropy profile as a function of $\epsilon$
at fixed $\lambda=0.46$. Low energy branches correspond to
double-cluster solution, the high-energy branch to the
``gas''-like equilibria. Notice that C is a maximum of
$S(\epsilon)$.}
\end{center}
\end{figure}

Nevertheless, the behaviour of the Boltzmann system beyond the
marginal stability line is quite intriguing. As we said, Eq.
(\ref{equa}) admits double-cluster type of solutions for all
$\epsilon$ if $\lambda>\lambda_c\simeq 0.45$, so there is no
Antonov limit. However, such solutions are locally unstable: there
exist perturbations of the density profile which lead to an
increase of entropy. A profile of the entropy density $S$ versus
$\epsilon$ at fixed $\lambda=0.46$ is shown in Fig. \ref{entr}. At
high energy the system is found in a stable, axially symmetric
state: a ``gas''. Upon decreasing $\epsilon$, the specific heat
$c_{\beta\omega}$ (\ref{cc}) stays positive until point C is
reached, where it turns negative (the curvature $\kappa_1$ of
$S(\epsilon,\lambda)$ or equivalently the quantity ${\rm
Hes}(\epsilon,\lambda)$ is zero). One enters the phase coexistence
region where collapsed configurations are possible. A further
(small) decrease in $\epsilon$ will drive us to the bifurcation
point D, where axial-rotational symmetry breaks down in favor of
such structures as ``binary stars''. (The axially symmetric
solution however persists, though locally unstable.) Decreasing
$\epsilon$ again one arrives at E, where double clusters become
locally unstable, that is a saddle point of $S[f]$. Double cluster
solutions (possibly more than one) however still exist left of E.
In such states the two ``stars'' are extremely dense and almost
point-like.

\section{Conclusion and outlook}

In summary, we have studied the equilibrium properties of a
self-gravitating and rotating gas assuming that particles obey
Boltzmann statistics in space. The physical properties of this
system can be understood easily as deriving from the balance of
(a) kinetic and gravitational energy at low angular momentum, and
(b) gravitational and rotational energy at high angular momenta.
In case (a), the effects of rotation are basically negligible
except for trivial distortions of the equilibrium shapes. If the
total energy is sufficiently high so that the kinetic term
dominates the gravitational one, the equilibrium state of the
system is axially-symmetric and gas-like. At low energies,
instead, gravitation induces the system to collapse into a dense
core in a low-density gas. In case (b), the high energy regime is
similar but at low energy rotation is strong enough to hinder the
collapse into a single cluster and drive the system to organize
into a double cluster structure (a ``binary star''). This simple,
intuitive picture is confirmed entirely by our mean-field theory,
which shows that in case (b) at least one main bifurcation (from
axially symmetric to axially asymmetric) in the solutions of the
basic Eq. (\ref{equa}) occurs, and possibly many others can be
found upon varying the conserved quantities $\epsilon$ and
$\lambda$.

Our work leaves several directions for further research. The main
question concerns probably the dynamics of these systems, which is
an equally old \cite{henon} and widely-discussed theme in
astrophysics, see e.g. \cite{lbe,what}. Among the many open issues
and possible extensions for the equilibrium case (such as the
effect of different masses, see \cite{inagaki,devega3}), we want
to put forward the following few. First is the behaviour of the
system left of the marginal stability line where double cluster
solutions become unstable. (We mentioned in Sec. 3 that the
location of this line turns out to depend on the maximum order of
harmonics included in the calculation of the gravitational
potential, $l_{\text{max}}$. Also the fact that no Antonov limit
exists for $\lambda>\lambda_c$ could be an effect of truncation.)
At low angular momenta, namely below Antonov limit, no statistical
equilibrium state exists and dynamical methods should be used to
grasp the physics of the system (see e.g. \cite{chavanis01} and
references therein). But for $\lambda>\lambda_c$ it is still
possible to obtain (unstable) equilibrium states as saddle point
of the entropy surface. This means that certain perturbations
applied to such ``singular'' double-cluster structures (highly
concentrated) will lead to an increase of entropy. It is not clear
to us whether local entropy maxima (metastable states) exist in
this region. Definitely, there are no global maxima for $S$ is
everywhere unbounded. This issue certainly deserves to be studied
further. Another important question that we didn't explore
concerns the relevance of mass as a further conserved quantity. If
$N$ is fixed, it should be treated on the same level as $\epsilon$
and $\lambda$, and this could lead to some qualitative changes in
the picture presented here. As a final consideration, we stress
the mean-field character of our solution. In spite of several
claims being made of mean-field theory being exact for
self-gravitating systems, we believe that it would be important to
investigate the effects of particle-particle correlations, also at
the merely quantitative level.

\medskip

\textbf{Acknowledgments.} Useful discussions with O. Fliegans are
gratefully acknowledged.

\end{document}